\documentclass[article]{elsarticle}
\usepackage{graphicx}
\usepackage{dcolumn}
\usepackage{bm}
\usepackage{amsmath}
\usepackage{amssymb}
\usepackage{lineno,hyperref}

\begin{document}
\begin{frontmatter}
\title{Fluctuations of Collective Coordinates and Convexity Theorems for
Energy Surfaces}

\author{B. G. Giraud}
\address{Institut de Physique Th\'eorique,
Centre d'Etudes Saclay, 91190 Gif-sur-Yvette, France}
\ead{bertrand.giraud@cea.fr}
\author{S. Karataglidis}
\address{Department of Physics, University of Johannesburg,
P. O. Box 524, Auckland Park, 2006, South Africa, and \\
School of Physics, University of Melbourne, Victoria, 3010, Australia}
\ead{stevenka@uj.ac.za}
\author{T. Sami}
\address{SUBATECH, UMR 6457, Universit\'e de Nantes, Ecole des Mines de
Nantes, IN2P3/CNRS, 4 rue Alfred Kastler, 44307 Nantes Cedex 3, France}
\ead{sami@subatech.in2p3.fr}

\begin{abstract}
Constrained energy minimizations of a many-body Hamiltonian return energy landscapes $e(b)$ where 
$b \equiv \langle B \rangle$ represents the average value(s) of one (or several)
collective operator(s), $B$, in an ``optimized'' trial state $\Phi_b$, and 
$e \equiv \langle H \rangle$ is the average value  of the Hamiltonian in this
state $\Phi_b$. It is natural to consider the uncertainty, $\Delta e$,
given that $\Phi_b$ usually belongs to a restricted
set of trial states. However, we demonstrate that the uncertainty, $\Delta b$,
must also be considered, acknowledging corrections to theoretical models. We also find 
a link between fluctuations of collective coordinates and convexity properties of
energy surfaces.
\end{abstract}
\date{\today}
\end{frontmatter}

\section{Introduction}
Collective coordinates \cite{BohMot} have been of central importance in
descriptions of structure and reactions in atomic, molecular, and nuclear physics. They generate models with
far less degrees of freedom than the true number of coordinates, $3 A$, as needed
for a microscopic description of a system of $A$ particles.  Often, the system's
dynamics can be compressed into slow motions of a few collective degrees
of freedom $B$, while the other, faster degrees can be averaged out. Also,
for identical particles, such collective degrees can be one-body
operators,
$B=\sum_{i=1}^A \beta(\mathbf{r}_i,\mathbf{p}_i,\sigma_i,\tau_i),$ where 
$\mathbf{r}_i, \mathbf{p}_i, \sigma_i, \tau_i$ refer to the position,
momentum, spin, and if necessary isospin, respectively, of particle $i$.
The summation over $i$ provides
for more inertia in $B$ than in the individual degrees $\beta_i$.

The concept of energy surfaces \cite{NixSwi} has been as important.
Given a ``coordinate-like'' collective operator $B$ and its expectation
value $b \equiv \left\langle B \right\rangle$, most collective models use
an energy function, $e(b)$, and also a $b$-dependent inertia parameter, that
drive the collective dynamics. Keywords such as ``saddles'', ``barriers'',
etc., flourish \cite{NixSwi}.

Simultaneously, it is often assumed that the function, $e(b),$ results
from an energy \textit{minimization} under constraint. Namely, while
the system evolves through various values of $b$, it is believed to tune its
energy to achieve a (local) minimum. This aspect of finding $e(b)$
is central to many fields of physics. To illustrate, consider a Hamiltonian,
$H=\sum_i T_i + \sum_{i<j} V_{ij},$ where $T$ and $V$ denote the
usual kinetic and interaction operators. Given a trial set of density
operators, $\mathcal{D}$, in many-body space, normalized by
$\text{Tr} \mathcal{D}=1$, the energy function $e(b)$ may be defined as,
\begin{equation}
e(b) = \inf_{\mathcal{D}\Rightarrow b} \text{Tr} \left\{ H \mathcal{D}
\right\},
\end{equation}
where $\text{Tr}$ is a trace in the many-body space for the $A$ particles.
The constraint, $\mathcal{D}\Rightarrow b,$ enforces $\text{Tr} \left\{ B
\mathcal{D} \right\}=b.$

There are theories which do not use, \textit{a priori}, an axiom of energy
minimization for the ``fast'' degrees of freedom. Time-dependent Hartree-Fock 
(HF) \cite{Bonche} trajectories, generalizations with pairing, adiabatic
versions \cite{Villars}, often show collective motions. Equations of motion \cite{Klein}
and/or a maximum decoupling \cite{Matsuyanagi} of 
``longitudinal'' from ``transverse''  degrees, have also shown significant
successes in the search for collective degrees, at the cost, however, of
imposing a one-body nature of both collective coordinates and 
momenta and accepting state-dependence of these  operators. Such approaches
define an energy surface once trajectories of wave functions have been
calculated. But they are not the subject of the present
analysis.  Herein, we focus on fixed operators constraining strict energy
minimizations within a fixed basis for single-particle and many-body states. The
questions arise: are those constraints themselves subject to fluctuations, and 
what would be the effect of those fluctuations on the energy minimization?

This is not a new question. The issue of constrained Hartree-Fock calculations was addressed
in Ref.~\cite{GLW}, which considered constrained Hartree-Fock calculations, and corrections to the energy surface.
 The issue has also been considered more recently in relation to high-energy
$(e,e'p)$ and $(p,2p)$ reactions \cite{Ry11}, where fluctuations in the position vectors of the
target nucleons involved in those processes were considered as going beyond the mean-field approximation
assumed for the structure of the target nucleus.

Ideally, to define mathematically a function $e(b)$ of the
collective coordinate, one should first diagonalize $B$ within the space
provided by the many-body states available for calculations \cite{Lesin}.
The resulting spectrum of $B$ should be continuous, or at least have a high
density for that chosen trial space. Then, for each eigenvalue $b$, one
should find the lowest  eigenvalue, $e(b)$, of the projection of $H$
into that eigensubspace labeled by $b$.

In practice, however, one settles for a diagonalization of the constrained
operator, $\mathcal{H} \equiv H - \lambda B$, where $\lambda$ is
a Lagrange multiplier,  or at least for a minimization of 
$\left\langle \mathcal{H} \right\rangle.$  Concomitantly, $B$ is assumed to 
have both upper and lower bounds, or that the constrained
Hamiltonian, $\mathcal{H},$ always has a ground state.
This returns the ``free energy'', 
$\varepsilon(\lambda) \equiv \left\langle \mathcal{H} \right\rangle$. The
label $b$ is no longer an eigenvalue but just an average value,
$b = \left\langle B \right\rangle$. A standard Legendre transform of 
$\varepsilon(\lambda)$ then yields the ``energy surface'', $e(b)$. This utilises the properties
$d\varepsilon/d\lambda=-\text{Tr}\{ B\mathcal{D} \} = -b$, and,
$de/db=\lambda$.
However we show in this work that constrained variation in a \textit{quantum}
system without additional precautions can raise at least two problems, namely:
i) the parameter $b$ may no longer be considered as a well-defined coordinate
for a collective model due to non-negligible fluctuations; we report
cases where the uncertainties, $\Delta b$, can vitiate the meanings of both $b$
and $e(b)$; and ii) there is a link between \textit{strict} minimization and the
curvature properties of $e(b)$ when fluctuations are at work.

Our argument is based fundamentally on a few theorems, presented in Section II, but also illustrated using
a few explicit models, which may be solved analytically or
numerically. Such solutions for those simple models are presented in Section III. A discussion and conclusion make
Section IV.

\section{Theorems linking strict minimization and convexity}

\subsection{Preliminaries}

Before proving such theorems, we must recall that, with Hartree-Fock (HF) and
Hartree-Bogoliubov (HB) \textit{approximations}, both convex and concave
branches are obtained for $e(b)$ by the addition, in 
$\langle \mathcal{H} \rangle$ with the constraint term
$-\lambda \left\langle B \right\rangle$, a square term,
$\frac{1}{2} \mu \left\langle B \right\rangle^2$, $\mu>0$ \cite{GLW,FQVVK,FQVVKa,SSBN}, allowing for
adjustable values of $\lambda$ and sometimes $\mu$. It must be stressed that
$\langle B \rangle^2$ differs from $\langle B^2 \rangle.$ It must also be
stressed that such mean-field results are approximations, and that all such
methods using a quadratic term in the original function, while
stabilizing the numerical procedure, amount to using an effective Lagrange
multiplier. Typically, when the ``completed'' functional, 
$\langle H \rangle + \frac{1}{2} \mu \langle B \rangle^2 - \lambda \langle B 
\rangle $, is minimized by a trial function $\phi$ in a mean-field
approximation, any bra variation, $\langle\delta \phi|$, induces the condition,
\begin{equation}
\left\langle \delta \phi \left| H \right| \phi \right\rangle + \left( \mu  \left\langle B \right\rangle - \lambda
\right) \left\langle \delta \phi \left| B \right| \phi \right\rangle = 0,
\label{effective}
\end{equation}
and a similar result holds for any ket variation. Clearly, this means that
the combination, 
$\langle H \rangle - \Lambda\,\langle  B \rangle$, with the effective Lagrange
multiplier, $\Lambda=\lambda - \mu\, \langle B \rangle$, has been made
stationary (but not necessarily minimal). 
We, therefore, consider the generic form, $\mathcal{H} = H - \lambda B$, in the 
following discussions.

As far as we are aware, none of the literature on energy surfaces obtained with
such quadratic cost functions is concerned with uncertainties in the collective
labels. However, any $e(b)$ may find an increase in its uncertainty if $b$ itself,
the expectation value of a quantum operator, is subject to large fluctuations. A surprising
observation of the present work is that the convexity properties of energy surfaces
and uncertainties in the collective coordinates are related. This work clarifies the
situation.

\subsection{Convexity from a convex domain of trial states}
Assume that minimization under constraint is performed in a \textit{convex}
domain of many-body density matrices $\mathcal{D}$. Namely, if $\mathcal{D}_1$
and $\mathcal{D}_2$ are trial states, any combination, 
$\mathcal{D}_m=\nu \mathcal{D}_1 + (1-\nu) \mathcal{D}_2$, with $0<\nu<1$, is
also a trial state. The constraint operator can be one bounded operator $B$,
or several such operators, $B_1, \dots, B_N$, or an infinite number of them,
such as, for density functionals, local field operators,
$\psi_r^{\dagger} \psi_r$, labeled by a continuous position $r$. For notational
simplicity, we denote the set of constraint operators by one symbol only, $B$,
and the corresponding expectation values as $b\equiv {\rm Tr} (B \mathcal{D})$.
We call ``slice $b$'' the set of all those $\mathcal{D}$ returning a given 
$b$. We also assume, for simplicity, that, \textit{inside} ``slice $b$'', the
minimum $e(b)$ of $\mathrm{Tr} (H \mathcal{D})$ is reached for a state 
$\mathcal{D}_{mb}$, which is non-degenerate.

Then, given two constraint values $b_1$ and $b_2$, with the associated
minimalizing states $\mathcal{D}_{m1}$ and $\mathcal{D}_{m2}$, it is clear
that the interpolating state, 
$\mathcal{D}_{mix}=\nu \mathcal{D}_{m1}+(1-\nu) \mathcal{D}_{m2}$,
returns both constraint and energy values, $b_{mix}=\nu b_1+(1-\nu) b_2$ and 
$e_{mix}=\nu e(b_1)+(1-\nu) e(b_2)$. However, there is no
reason why $\mathcal{D}_{mix}$ should be the energy minimizing state inside
that slice labeled by $b_{mix}$. Hence, by necessity, the energy minimum inside
the slice obeys the inequality, $e(b_{mix}) \le e_{mix}$. In short, $e$ is 
necessarily a convex function (or functional).

Experimental tables of ground state energies $E_A$ are incompatible (see 
Ref. \cite{BG}) with such a convexity if a density functional is requested to 
be universal in terms of a particle number $A$. Indeed, there are many cases
where $E_A \ge (E_{A-1}+E_{A+1})/2$.
In such a situation, a convex functional, with a mixed density, 
$(\rho_{A-1}+\rho_{A+1})/2$, hence a fluctuation $\Delta A =1$, will return a 
lower energy than $E_A$. See Ref. \cite{BG} for ways to ``make Nature convex''.

Many practical calculations of spectra or energy surfaces use at first mean
field methods, where trial states do not belong to a convex domain. For
instance, a weighted sum of two Slater determinants usually does not make
a Slater determinant. A similar statement holds for HB states. The question of
convexity \cite{NPF}, however, and its relation to fluctuations, often
remains.

\subsection{A general theorem}

Given a constrained Hamiltonian, $H-\lambda B$,
consider a solution branch $\mathcal{D}(\lambda)$, expanding up to second
order, assuming that the manifold of solutions is suitably analytic,
\begin{equation}
\mathcal{D}(\lambda+d \lambda) = \mathcal{D}(\lambda)+d\lambda
 (d\mathcal{D}/d\lambda) + (d\lambda^2/2) (d^2\mathcal{D}/d\lambda^2).
\label{nalytic2}
\end{equation}
In those cases where pure states, ${\cal D}=|\phi\rangle \langle \phi|$, are
used, the wave function of interest is assumed to be analytical, and, clearly,
$d {\cal D} / d \lambda = | \phi \rangle \langle d \phi / d \lambda | +
|d \phi / d \lambda \rangle \langle \phi |$, and also,
$d^2 {\cal D} / d \lambda^2= | \phi \rangle \langle d^2 \phi / d \lambda^2 | +
2\, | d \phi / d \lambda \rangle \langle d \phi / d \lambda| +
| d^2 \phi / d \lambda^2 \rangle \langle \phi |$.

The stationarity and minimality of 
$\text{Tr}\left\{ \mathcal{H} \mathcal{D} \right\}$ with respect to any
variation of $\mathcal{D},$ and in particular with respect to that variation,
$\mathcal{D}(\lambda+d\lambda)-\mathcal{D}(\lambda),$ induce,
\begin{align}
\text{Tr} \left\{ \mathcal{H} d\mathcal{D}/d \lambda\right\} & = 0, \nonumber
\\ 
\text{Tr} \left\{ \mathcal{H}d^2\mathcal{D}/d \lambda^2 \right \} & \ge 0.
\label{gensolcond}
\end{align}
The free energy $\varepsilon$ is also stationary for 
$\mathcal{D}(\lambda + d\lambda)$, but the Hamiltonian is now, 
$\mathcal{H}(\lambda) - B\, d\lambda$, and the derivative of the state is,
$d\mathcal{D}/d\lambda + d\lambda (d^2\mathcal{D}/d\lambda^2) +
\mathcal{O}(d\lambda^2)$, hence,
\begin{equation}
\text{Tr} \left\{ \left( \mathcal{H} - Bd\lambda\right) 
\left[ d\mathcal{D}/d\lambda + 
d\lambda (d^2\mathcal{D}/d\lambda^2)+\mathcal{O}(d\lambda^2) \right] \right\} 
= 0.
\label{genstatio}
\end{equation}
The zeroth order of this, Eq.~(\ref{genstatio}), is, 
$\text{Tr} \left\{\mathcal{H} d\mathcal{D}/d\lambda \right\}$. It vanishes,
because of the first of Eqs.~(\ref{gensolcond}). The first order, once
divided by $d \lambda$, gives,
\begin{equation}
- \text{Tr} \left\{B d\mathcal{D}/d\lambda \right\} =
- \text{Tr} \left\{\mathcal{H} d^2\mathcal{D}/d\lambda^2 \right\}.
\label{trick}
\end{equation}
The left-hand side of Eq.~(\ref{trick}) is nothing but the second 
derivative, $d^2 \varepsilon/d\lambda^2$. The right-hand side is
semi-negative-definite, because of the second of Eqs.~(\ref{gensolcond}).
Hence, the plot of $\varepsilon(\lambda)$ is a concave curve and the plot
of its Legendre transform, $e(b)$, is convex. (Other papers \cite{BG}
have the opposite sign convention of the second derivative to define
convexity.) With the present sign convention, strict minimization of 
$\langle {\cal H} \rangle$ necessarily induces convexity of $e(b)$.
Any concave branch for $e(b)$ demands an explanation.

Note that this proof does not assume any specification of 
$\mathcal{D}(\lambda)$, whether it is constructed either from exact or
approximate eigenstates of $\mathcal{H}$. Therefore strict minimization of
$\langle (H-\lambda B)\rangle$ can \textit{only} return \textit{convex}
functions $e(b)$. Maxima are impossible. In the generalization where
several collective operators $B_1, \dots, B_N$, are involved, convexity
still holds, so saddles are also excluded. Hence, only an absolute minimum
is possible. (However, we shall recall below how to overcome the paradox: by
\textit{keeping small enough} \cite{PhysScrip} \textit{the fluctuations} of the
collective coordinate(s), one can deviate from convexity, and more important,
validate the quality of the representation provided by branches
 $\mathcal{D}(\lambda)$.)

\subsection{Same theorem, for exact solutions}

Let  $\psi(\lambda)$ be the ground state of $\mathcal{H}$. (For the sake of
simplicity, we assume that there is no degeneracy.) The corresponding
eigenvalue, $\varepsilon(\lambda)$, is stationary with respect to variations
of $\psi$, among which is the ``on line'' variation,
$d\lambda\, (d\psi/d\lambda)$, leading to the well-known first derivative,
$d\varepsilon/d\lambda = -b \equiv - \left\langle \psi \left| B \right|\psi 
\right\rangle$.
Consider the projector,
$Q = 1- \left| \psi \right\rangle \left\langle \psi \right|.$
Brillouin-Wigner theory yields the first derivative of $\psi$, \textit{viz.}
\begin{equation}
\frac{d\left|\psi \right\rangle}{d\lambda} =
- \frac{Q}{\varepsilon-Q\mathcal{H}Q} B \left| \psi \right\rangle.
\end{equation}
This provides the second derivative of $\varepsilon,$
\begin{equation}
-\frac{d b}{d \lambda} \equiv - \frac{d}{d \lambda} \left\langle \psi \left|
 B \right| \psi \right\rangle = 2 \left\langle \psi \left| B 
\frac{Q}{\varepsilon-Q\mathcal{H}Q} B \right| \psi \right\rangle.
\label{negatdefin}
\end{equation}
Since the operator $(\varepsilon-Q\mathcal{H}Q)$ is clearly
negative-definite, the eigenvalue, $\varepsilon$, is a concave function of
$\lambda$. It is trivial to prove that the same concavity holds for the
ground state eigenvalue $\varepsilon(\lambda_1, \dots,\lambda_N)$ if several
constraints, $B_1, \dots, B_N$, are used. If, moreover, a temperature 
$T$ is introduced, the thermal state, 
$\mathcal{D}=\exp\left[-\mathcal{H}/T\right] / 
\text{Tr}\exp\left[- \mathcal{H}/T\right],$
replaces the ground state projector and
the free energy
also contains the entropy contribution, $-T S,$ where
$S=-\text{Tr} \left\{ \mathcal{D} \ln \mathcal{D} \right\}$. A proof of the
concavity of the exact $\varepsilon(\lambda_1, \dots,\lambda_N;T)$ is also
easy \cite{Balian}.

At $T=0,$ the usual Legendre transform expresses the energy,
$e \equiv \left\langle \psi \left| H \right| \psi \right\rangle$, in terms
of the constraint values, $b_1,\dots,b_N$, rather than the Lagrange
multipliers. For simplicity, consider one constraint only; the generalization
to $N>1$ is easy. Since $e \equiv \varepsilon + \lambda b$, then
$d e/d b = \lambda$, a familiar result for conjugate variables.
Furthermore, the second derivative, $d^2 e/d b^2,$ reads, 
$d\lambda/db = 1/(db/d\lambda)$. From Eq.~(\ref{negatdefin}), the derivative,
$db/d\lambda$, is positive-definite. Accordingly, $e$ is a convex function
of $b$. Now, if $T > 0$, the Legendre transform instead generates a reduced
free energy, $\eta \equiv (e - TS)$, a convex function of the constraint
value(s). An additional Legendre transform  returns $e$ alone, as a convex
function of the constraint(s) and $S$.

Let $b_-$ and $b_+$ be the lowest and highest eigenvalues of $B.$ When
$\lambda$ runs from $-\infty$ to $+\infty,$  then $b$ spans the interval,
$[b_-,b_+]$. There is no room for junctions of convex and concave branches
under technical modifications as used by Refs. \cite{GLW,FQVVK,FQVVKa,SSBN}.
For every exact diagonalization of $\mathcal{H}$, or exact partition function,
convexity sets a one-to-one mapping between $b$ in this interval and $\lambda$.
More generally, with exact calculations, there is a one-to-one mapping
between the set of Lagrange multipliers,
$\left\{\lambda_1, \dots,\lambda_N \right\},$
and that of obtained values, $\left\{ b_1, \dots, b_N \right\}$, of the
constraints. Convexity, \textit{in the whole obtained domain of constraint
values}, imposes a poor landscape: there is one valley only.

\begin{figure}
\centering\scalebox{0.68}{\includegraphics*{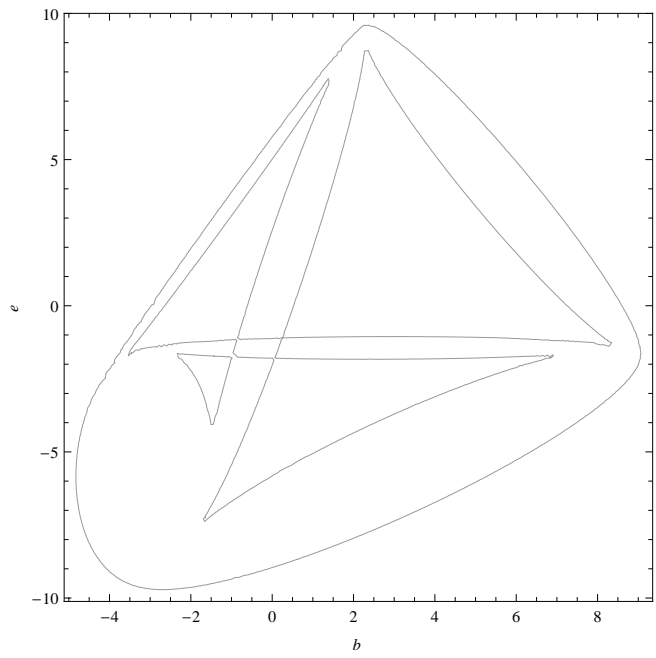}}
\caption{\label{figure5} Branches obtained for $e(b)$ by algebraic elimination
of the Lagrange multiplier $\lambda$ between $b(\lambda)$ and $e(\lambda)$ 
in a toy case with $4\times4$ matrices. Convexity of the ground state branch.}
\end{figure}

We tested this surprising result with several dozens of numerical cases, 
where we used random matrices for $H$ and $B,$ with various dimensions. As
an obvious precaution, we eliminated those very rare cases where both $H$ and
$B$ turned out, by chance, to be block matrices with the same block structure;
such cases give rise to level crossings and degeneracies. Then every remaining 
situation, without exception, confirmed the convexity of $e$. 
Figure \ref{figure5} shows all branches of $e(b)$ for  
$H=\left[ \begin{smallmatrix}
 -3 & 0 & 3 & 2 \\ 0 & 5 & -4 & 4 \\ 3 & -4 & -4 & 5 \\ 2 & 4 & -5 & -1
\end{smallmatrix} \right]$ 
and 
$B=\left[ \begin{smallmatrix}
 0 & -2 & 0 & 3 \\ -2 & 2 & -2 & -2 \\ 0 & -2 & -4 & -1 \\ 3 & -2 & -1 & 7
\end{smallmatrix} \right]$
for instance. Such branches are easily derived algebraically \cite{PLB} from
the polynomial, $P \equiv \det(H-\lambda B-\varepsilon)$. The convexity of
the ground state branch is transparent. Clear also are its infinite derivatives
when $b$ reaches $b_-=-4.83$ and $b_+=9.60$, and the vanishing derivative at
the point corresponding to the unconstrained ground state, where $b=-2.69$ and
$e=e_-=-9.71$.

\subsection{Same theorem, for approximations via constrained HF calculations}

Consider now energy surfaces obtained from approximations. Typically,
one uses a HF or HB calculation, at zero or finite $T$. Trial states in
such methods span a nonlinear manifold; indeed, a sum of two determinants
is usually not a determinant. Let $\mathcal{D}(\lambda)$ denote one
$A$-body density operator where, within such nonlinear approximations, a
minimum, $\varepsilon(\lambda)$, of
$\text{Tr} \left\{\mathcal{H} \mathcal{D}\right\}$ or of
$\left(\text{Tr} \left\{\mathcal{H} \mathcal{D}\right\}-TS\right)$, is
reached. It may be degenerate, but, in any case, it is stationary for
arbitrary variations $\delta \mathcal{D}$ within the set of trial states.
Accordingly, the first derivative again reads,
$d \varepsilon/d \lambda = -b \equiv - \text{Tr} \left\{ B
 \mathcal{D}\right\}$. Then, if a Legendre transform holds, defining
$\eta \equiv \varepsilon + \lambda b$ in terms of $b$, the same argument that
was used for the exact case again yields, $d \eta/d b = \lambda$. With
$N$ constraints, the gradient of $\eta$ in the domain spanned by
$\left\{b_1,\dots,b_N \right\}$ is the vector $\left\{\lambda_1,\dots,
\lambda_N \right\}$.

To discuss second derivatives, consider, for instance,
HF calculations, where $A$-body density operators are dyadics of determinants,
$\mathcal{D}= \left| \phi \right\rangle \left\langle \phi \right|$.
Norm-conserving variations of an HF solution, $\phi$, can be parametrized
as, $\left| \delta \phi \right\rangle = \exp(i X \delta \alpha ) \left| 
 \phi \right\rangle,$  with $X$ an arbitrary particle-hole Hermitian
operator, and $\delta \alpha$ an infinitesimal coefficient. Under such a
variation in the neighborhood of a HF solution, the first and second order
variations of the free energy,
$\varepsilon \equiv \text{Tr} \left\{ \mathcal{H} \exp(i X \delta \alpha ) 
\mathcal{D} \exp(-i X \delta \alpha ) \right\}$, read,
\begin{equation}
\delta \varepsilon=i \delta \alpha \text{Tr} \left\{ \left[\mathcal{H},X\right]
\mathcal{D} \right\} = 0,
\label{statio}
\end{equation}
and
\begin{equation}
\delta^2 \varepsilon=-\left(\delta \alpha^2/2\right) \text{Tr} 
\left\{ \left[\left[ \mathcal{H},X\right],X\right]\mathcal{D} \right\} \ge 0,
\label{mini}
\end{equation}
respectively. If $\mathcal{D}$ is a HF solution, the first order
vanishes $\forall X.$ Since only those solutions that give
minima are retained, the second order variation of $\varepsilon$ is 
semi-positive-definite, $\forall X$ again. Now, when $\mathcal{H}$ receives
the variation, $- B d \lambda$, there exists a particle-hole operator, $Y$, a
special value of $X$, that, with a coefficient $d\lambda$, modifies the
solution. This reads
$\left| \Phi \right\rangle = \exp(i  Y d\lambda ) \left|\phi \right\rangle.$
Simultaneously,
those particle-hole operators that refer to this new Slater determinant 
$\Phi$ become $\mathcal{X}=\exp(i Y d\lambda) X \exp(-i Y d\lambda)$. 
The new energy is, 
\begin{equation}
\varepsilon_Y = \text{Tr}\left\{\exp(-i Y d\lambda ) (\mathcal{H} - B d\lambda)
\exp(i Y d\lambda )\mathcal{D} \right\}. 
\end{equation}
The stationarity condition, Eq. (\ref{statio}), becomes,
\begin{align}
0 & = \text{Tr}\left\{\exp(-i Y d\lambda)\left[ (\mathcal{H} - Bd\lambda),
\mathcal{X}\right] \exp(i Yd\lambda) \mathcal{D}\right\} \nonumber \\
& = \text{Tr} \left\{\left[ \exp(-i Y d\lambda)(\mathcal{H} - B d \lambda)
\exp(i Yd\lambda) , X \right] \mathcal{D} \right\}.
\label{newstatio}
\end{align}
The zeroth order in $d\lambda$ of this, Eq.~(\ref{newstatio}), reads,
$\text{Tr}\left\{\left[ \mathcal{H}, X \right]\mathcal{D}\right\}$, and
vanishes $\forall X$, because of Eq.~(\ref{statio}). Then the first order in
$d\lambda$ gives, again $\forall X$,
\begin{equation}
\text{Tr} \left\{\left[ B,X \right]\mathcal{D}\right\} = 
i\text{Tr} \left\{\left[ \left[ \mathcal{H} , Y \right] , X \right]
\mathcal{D}\right\}.
\label{astuce}
\end{equation}
The second derivative is,
\begin{align}
d^2\varepsilon/d\lambda^2 & =-(d/d \lambda) \text{Tr} \left\{
\exp(-i Y d\lambda) B \exp(i Yd\lambda)\mathcal{D} \right\}\nonumber \\ 
&  = -i \text{Tr} \left\{ \left[ B,Y \right] \mathcal{D} \right\}.
\end{align}
Upon taking advantage of Eq.~(\ref{astuce}), for $Y$ as a special case
of $X,$ this becomes,
\begin{equation}
d^2\varepsilon/d\lambda^2 =
\text{Tr} \left\{\left[\left[ \mathcal{H},Y \right],Y\right ] \mathcal{D}
\right\},
\end{equation}
the right-hand side of which is semi-negative-definite, see Eq.~(\ref{mini}).
The solution branch obtained when $\lambda$ runs is, therefore, concave. Its
Legendre transform is convex.

\subsection{About concave branches}

Let us return to the ``quadratic cost'' function, that redefines the 
``free energy'' to be minimized, $\mathcal{E}$, as,
$\mathcal{E}=\langle H \rangle + \frac{1}{2} \mu \langle B \rangle^2 - 
\lambda \langle B \rangle$, or, in a shorter notation, 
$\mathcal{E}=e+\frac{1}{2}\mu b^2-\lambda b$.  Let the positive number $\mu$
be kept constant, allowing $\lambda$ to vary. It is easy to show again that 
$d \mathcal{E}/d \lambda=-b$.
But now one finds that, $de/db=\Lambda \equiv \lambda -\mu b$, where the
effective Lagrange multiplier $\Lambda$ is that noticed at the stage
of Eq. (\ref{effective}).
Assume again that variational solutions are analytic with respect to 
$\lambda$. For the sake of simplicity, use wave functions $\phi$ rather than
density operators, hence $\mathcal {D}=|\phi \rangle \langle \phi|$. 
The analog of  Eq. (\ref{nalytic2}) reads, 
$\phi(\lambda+d\lambda)=\phi+d\lambda\, \phi' + \frac{1}{2} (d\lambda)^2 \, 
\phi''$, with short notations for derivatives of $\phi$. The stationarity of
$\mathcal{E}$ then reads, 
\begin{equation}
0 = \mathcal{E}' \equiv \langle \phi' |(H-\Lambda B)| \phi \rangle + 
\langle \phi |(H-\Lambda B)| \phi' \rangle\, .
\label{initstat} 
\end{equation}
If this ``quadratic cost'' $\mathcal{E}$ is a strict minimum with respect
to variations of $\phi$, one finds,
\begin{align}
0 < \mathcal{E}'' \equiv \langle \phi'' |\, (H-\Lambda B)\, | \phi   \rangle + 
   2 \langle \phi'  |\, (H-\Lambda B)\, | \phi'  \rangle \nonumber \\
 +\, \langle \phi   |\, (H-\Lambda B)\, | \phi'' \rangle + \mu \, 
(\, \langle \phi' |B| \phi \rangle + \langle \phi |B| \phi' \rangle\, )^2\, .
\label{strict}
\end{align}
When $\lambda$ becomes $\lambda+d\lambda$, the derivative of the wave
function becomes $(\phi'+d\lambda\, \phi'')$ and the functional to be
minimized receives an additional contribution,
$-d\lambda\, \langle B \rangle$. Then the stationarity condition at 
$\phi(\lambda+d\lambda)$ becomes,
\begin{equation}
0=(\mathcal{E}'+d\lambda\, \mathcal{E}'')-d\lambda\,
 (\, \langle \phi' |B| \phi \rangle + \langle \phi |B| \phi' \rangle\, )\, .
\end{equation}
This simplifies into, $\mathcal{E}''=db/d\lambda$, since $\mathcal{E}'=0$,
see Eq. (\ref{initstat}), and since, simultaneously,
$db/d\lambda=\langle \phi' |B| \phi \rangle + \langle \phi |B| \phi' \rangle$.
Furthermore, since $\mathcal{E}''>0$, see Eq. (\ref{strict}), and since we
know that $d\mathcal{E}/d\lambda=-b$, we find that 
$d^2\mathcal{E}/d\lambda^2=-\mathcal{E}''$, a negative quantity. The function,
$\mathcal{E}(\lambda)$, is, therefore, necessarily concave if a strict
minimization of $\mathcal{E}$ has been performed in the domain of trial 
states.

As stated at the beginning of this Subsection, the ``energy surface'' 
function, $e(b)$, now obeys the condition, $de/db=\lambda-\mu\, b$, hence 
$d^2 e/db^2=d\lambda/db-\mu$. While we found that 
$d\lambda/db=1/(db/d\lambda)=1/\mathcal{E}''$ is a positive quantity, the 
term, $-\mu$, competes with $d\lambda/db$ and may induce negative values of
$d^2 e /db^2$, hence concave regions in the plot of $e(b)$, where, 
incidentally, the second derivative, $d^2e/db^2$, cannot be more negative 
than $-\mu$.

Consider two solutions, $\phi_1$, $\phi_2$, generating two points, $(b_1,e_1)$,
$(b_2,e_2)$, that are separated by an inflection point of $e(b)$ and that
show the same derivative, $\Lambda=(de/db)_1=(de/db)_2$, in other words,
$\lambda_1 - \mu b_1 = \lambda_2 - \mu b_2.$ Denote $\phi_1$ the lower energy
solution, hence $e_1<e_2$. Then $\phi_2$, compared with $\phi_1$, is an
excited solution of the variational stationarity problem for the same
constrained Hamiltonian, $H-\Lambda B$, that drives $\phi_1$. For the sake of
the argument, assume that $b_1<b_2$. Then along the convex branch of $e(b)$,
the value of $\Lambda$ increases when $b$ increases, and, when $b$ decreases
along the concave branch, the value of $\Lambda$ also increases. When one
reaches the inflection point, $(b_i,e_i)$, a maximum value, $\Lambda_i$,
is reached. One must conclude that the two, formerly distinct solutions,
$\phi_1(\Lambda)$ and $\phi_2(\Lambda)$, do not cross and, rather, smoothly
fuse into a unique solution $\phi(\Lambda_i)$. This is a very unlikely 
situation for exact solutions, namely eigenstates, as is well known. Only 
approximations can afford such an anomaly. (Notice also that, because 
$H-\Lambda B$ is assumed bounded from below, $\forall \Lambda$, there will
exist a stationary solution $\phi_+$ for any multiplier 
$\Lambda_+ > \Lambda_i$, even infinitesimally close to $\Lambda_i$.
Then one must accept that $\Phi_+$ and $\Phi_i$ strongly differ from each
other.) 

The zoo of \textit{stationary} solutions of approximate methods such as mean
field methods  (non linear!) can be rich enough to accommodate such
singularities. This makes a paradox: would non-linear approximations
generate a more flexible, physical tool than exact solutions?
``Phase transitions'', a somewhat incorrect wording for a finite system,
are sometimes advocated to accept continuing branches of energy minima
into metastable branches. But this definitely claims some caution with
the axiom of strict energy minimization to freeze fast degrees. This need for
excited solutions in constrained mean field calculations, namely two
solutions for the same value of a Lagrange multiplier to describe both
sides of an inflection point of the barrier, is well known and used. See in
particular \cite{FQVVK,FQVVKa}, where a tangent parabola rather than a tangent
straight line is used to explore an energy surface by mean field methods.

\subsection{Bimodal solutions in mean field approximations}

Besides the caution about the ``fast degree minimization hypothesis'' one
should consider whether such mean field solutions, in convex or
concave branches, might be vitiated by large uncertainties for $b$. The 
following solvable models give a preliminary answer.

Consider $N$ identical, 1-D fermions with Hamiltonian,
\begin{equation}
H=\sum_{i=1}^N p_i^2/(2m)+M \omega^2 R^2/2+\sum_{i>j=1}^N v_{ij},
\label{bareH}
\end{equation}
where $R=\sum_i r_i/N$ is the center-of-mass (c.m.) position, $p_i,r_i,m$
denote the single particle momentum, position and mass, respectively, of each
fermion, and $M=N m$ is the total mass. The c.m. momentum is, $P=\sum_i p_i.$
We use a system of units such that $\hbar=m=\omega=1,$ where $\omega$
denotes the frequency of the c.m. harmonic trap. The interaction, $v,$ is set
as Galilean invariant and so is the sum, $V=\sum_{i>j} v_{ij}.$ In the 
following, $v$ is taken as a spin and isospin independent and local force,
$v_{ij}=v(|r_i-r_j|).$

The collective operator we choose to constrain $H$ is a half sum of 
``inertia'' (mass weighted square radii), $B=\sum_{i=1}^N m r_i^2/2$. The
constrained Hamiltonian then reads,
\begin{equation}
{\cal H}=\sum_i p_i^2/(2m)+M \omega^2 R^2/2+\sum_{i>j} v_{ij}-\lambda
\sum_i m r_i^2/2.
\label{cnstHH}
\end{equation}

Let $\xi_1=r_2-r_1,$ $\xi_2=r_3-(r_1+r_2)/2,$ ..., 
$\xi_{N-1}=r_N-(r_1+r_2+\dots+r_{N-1})/(N-1)$ denote the usual Jacobi 
coordinates with $\Pi_{\alpha}$ and $\mu_{\alpha}$, $\alpha=1,\dots,(N-1)$,
the corresponding momenta and reduced masses. In this Jacobi
representation the constraint becomes, 
$B=M R^2/2+\sum_{\alpha} \mu_{\alpha} \xi_{\alpha}^2/2.$ Accordingly, the 
constrained Hamiltonian decouples as a sum of a c.m. harmonic oscillator,
\begin{equation}
\mathcal{H}_{\text{c.m.}} = \frac{P^2}{(2M)} + \frac{1}{2}M \Omega^2 R^2,
\; \; \; \Omega^2=\omega^2-\lambda,
\label{renortrap}
\end{equation}
provided  $\lambda<\omega^2$, and an internal operator,
\begin{equation}
\mathcal{H}_{\text{int}}=\sum_{\alpha} \left[ \Pi_{\alpha}^2/(2 \mu_{\alpha}) 
- \lambda \mu_{\alpha} \xi_{\alpha}^2/2 \right]+ V.
\label{intern}
\end{equation}
With the present power of computers and present experience with
Faddeev(-Yakubovsky) equations, this choice of $\mathcal{H}$ and $B$, with
its ability to decouple, provides soluble models with, typically,
$N=2,3,4$. (Decoupling also occurs if $B$ is a quadrupole operator.) 
Exact solutions can thus be compared with mean field approximations
and validate, or invalidate, the latter. Here, however, we are not
interested in the comparison, but just in properties of the mean 
field solutions as regards $B$.

For this, as long as $N$ does not exceed $4$, we assume that the
Pauli principle is taken care of by spins and isospins, understood in the
following, and that the space part of the mean field approximation is a
product, $\phi=\prod_i \varphi(r_i),$ of identical, real and positive parity
orbitals. The corresponding Hartree equation reads,
\begin{equation}
\left[\frac{p^2}{2m} + \Lambda r^2 + u(r) \right] \varphi(r)= \varepsilon_{sp}
\varphi(r),
\end{equation}
with $u(r)=(N-1) \int_{-\infty}^{\infty} v(r-s) [\varphi(s)]^2$ and 
$\Lambda=M \omega^2/(2 N^2)-\lambda m/2.$ The term,
$M \omega^2/(2 N^2),$ clearly comes from the c.m. trap. The same trap induces
a two-body operator, $M \omega^2 \sum_{i \ne j} r_i r_j/(2 N^2),$ which cannot
contribute to the Hartree potential, $u,$ since every dipole moment, 
$\left\langle r_j \right\rangle$, identically vanishes here.

Once $\varphi$ and $\varepsilon_{sp}$ are found, one obtains the free energy, 
$\varepsilon_{\text{Hart}}=\langle \mathcal{H} \rangle=N \varepsilon_{sp}-
N \left\langle 
\varphi \left| u \right| \varphi \right\rangle/2$, then the value of the
constraint, 
$\left\langle B \right\rangle_{\text{Hart}} = N m \left\langle \varphi 
\left| r^2 \right| \varphi \right\rangle/2,$ and
the square fluctuation, $(\Delta b)^2_{\text{Hart}} = N m^2 
\left( \left\langle \varphi \left|
 r^4 \right| \varphi \right\rangle - \left\langle \varphi \left| r^2 \right| 
\varphi \right\rangle^2 \right)/4.$
The physical energy, $e_{\text{Hart}}(b),$ in this Hartree approximation,
clearly obtains by adding $\lambda \langle B \rangle_{\text{Hart}}$ to 
$\varepsilon_{\text{Hart}}.$

\begin{figure}
\centering\scalebox{0.68}{\includegraphics*{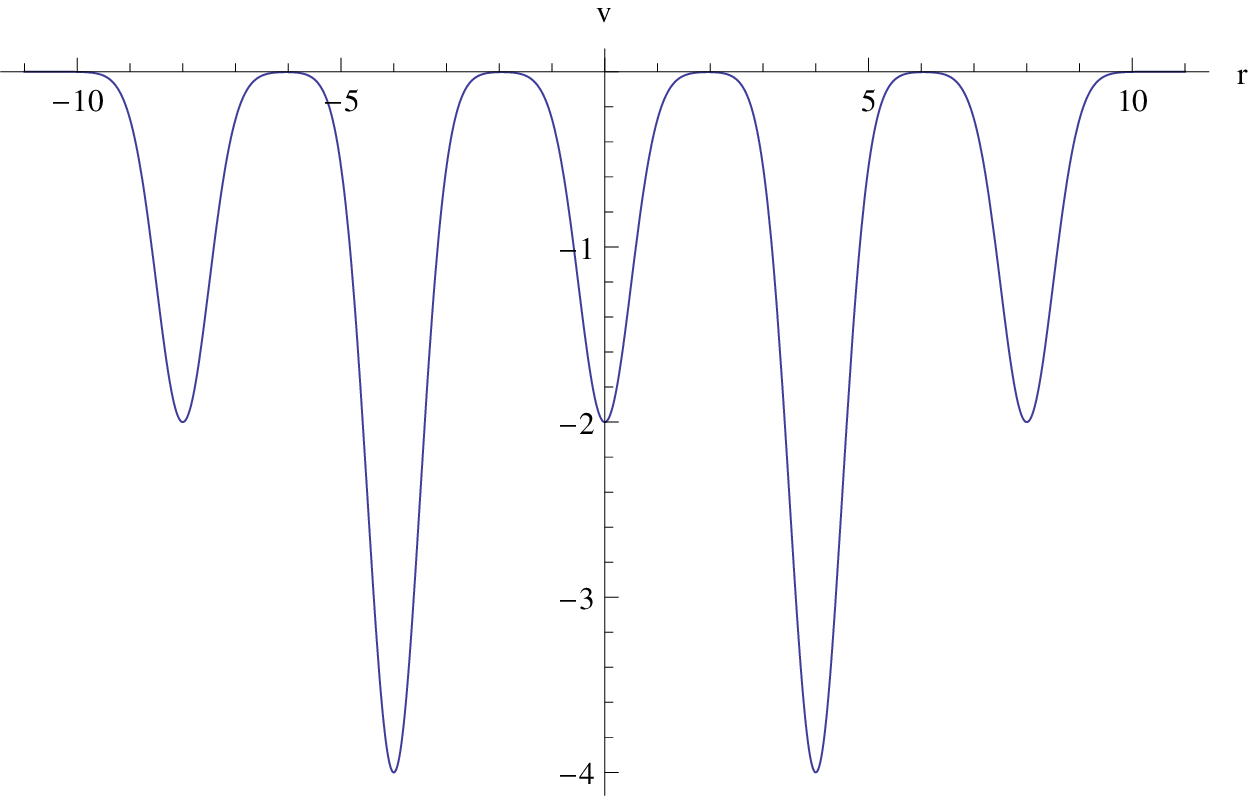}}
\caption{\label{figure6} An interaction giving multimodal Hartree solutions.}
\end{figure}

We show now, among many cases we studied, Hartree results when 
$v_{ij}=-2 [\exp(-2 (r_i-r_j+8)^2)  +  \exp(-2 (r_i-r_j-8)^2) +
          2 \exp(-2 (r_i-r_j+4)^2) + 2 \exp(-2 (r_i-r_j-4)^2) + 
            \exp(-2 (r_i-r_j)^2)],$ see Fig. \ref{figure6}.

\begin{figure}
\centering\scalebox{0.68}{\includegraphics*{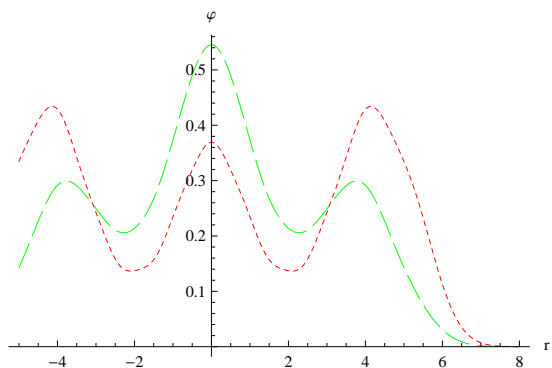}}
\caption{\label{figure7} Dashes, Hartree orbital $\varphi(r)$ if $N=2,$  
$\lambda=.47$ for the interaction shown in Fig. \ref{figure6}. Dots, the 
same for $\lambda=.55.$}
\end{figure}

\begin{figure}
\centering\scalebox{0.68}{\includegraphics*{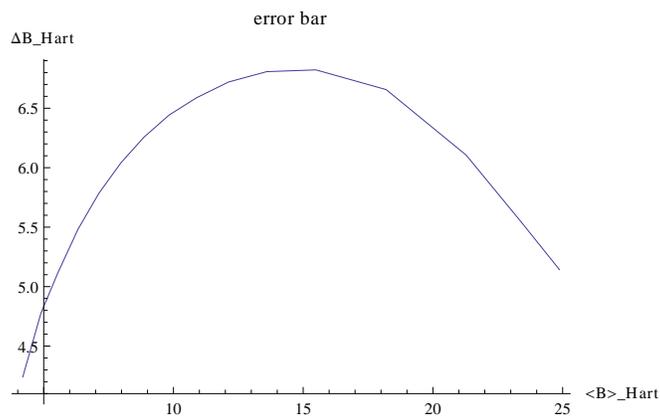}}
\caption{\label{figure8} Same Hartree model. $\Delta b$ as a function
of $\langle B \rangle.$}
\end{figure}

The results shown in Figs. \ref{figure6}-\ref{figure8} correspond to $N=2$. 
Similar results hold with $N=3,4$. Orbitals $\varphi$ are expanded in the
first 11 even states of the standard harmonic oscillator. The only difference 
between the two orbitals shown in Fig. \ref{figure7} is the value of 
$\lambda$. Both orbitals, and many other ones, when $\lambda$ runs, show a 
bimodal structure, their left-hand-side and right-hand-side bumps being 
equivalent with respect to the even observable, $B \propto r^2$. Whether
$N=2$, $3$, or $4$, we found many cases where the value of $\langle B \rangle$
has nothing to do with the positions of the peaks of $\varphi$. The bad
quality induced by the corresponding uncertainties is illustrated by 
Fig. \ref{figure8}. Therefore, little trust is available for the energy curve,
$e_{Hart}(b)$, that results from this model.

\section{Models showing the existence and extent of quantum
fluctuations of collective coordinates}

We now turn our attention to illustrating the effects of fluctuations of
collective coordinates, as discussed above, with various examples.

\subsection{Fluctuation of the mean square radius in a harmonic oscillator
shell model}

Consider $A$ fermions with Hamiltonian,
$H=\frac{1}{2} \sum_{i=1}^A (p_{xi}^2+p_{yi}^2+p_{zi}^2+x_i^2+y_i^2+z_i^2)$,
or, in a shorter notation, $H=\sum_i (p_i^2+r_i^2)/2$. Ignore spin (and
isospin) and fill completely each shell up to that one with energy 
$n+\frac{3}{2}$. It is trivial that the sequence of individual shell
populations reads, $1,3,6,\dots,(n+1)(n+2)/2$, and that, when all shells up
to and including  ``shell $n$'' are filled, the particle number reads,
$A=(n+1)(n+2)(n+3)/6$. Denote $\phi$ that Slater determinant made of such
$(n+1)$ filled shells and choose now the collective operator,
$B=A^{-1} \sum_{i=1}^A r_i^2$. By definition, $b\equiv \langle B \rangle$
represents the square of an average radius, $\bar r=\sqrt{b}$.

For the sake of mathematical rigour, it might be necessary to redefine $B$ with
a cut-off, such as, $B_c=\sum r_i^2 \exp(-r_i^2/R_c^2)$, with $R_c$ 
significantly larger than atomic, molecular or nuclear radii, to ensure that
constrained Hamiltonians, $H-\lambda B_c$, do have a ground state if 
$\lambda>0$. But this technicality can be neglected in practice, with 
calculations using a large but finite basis of finite range variational
states. 

Each shell with index $m,\ m=0,\dots,n $, contributes 
$A^{-1} (m+3/2) (m+1)(m+2)/2$ to $\langle \phi | B | \phi \rangle$, hence a 
total $\langle B \rangle = A^{-1} (n+1) (n+2)^2 (n+3)/8$.
Consider now $\langle B^2 \rangle$, to calculate the fluctuation.
Let $h$ be those states filled in $\phi$ and $P$ the other
eigenstates of $H$. The one-body operator $B$ can excite at most 
one-particle-one-hole states when acting upon $\phi$, hence,
\begin{equation}
\langle B^2 \rangle=\langle \phi|B|\phi \rangle \langle \phi|B|\phi
\rangle + \sum_{Ph} \langle \phi|B|\phi_{Ph} \rangle \langle \phi_{Ph}|B|\phi
\rangle\, .
\end{equation}
Accordingly, $(\Delta b)^2$ reduces to a sum of particle-hole squared
matrix elements, $(\Delta b)^2=A^{-2} \sum_{Ph} (\langle P |r^2| h \rangle)^2$.
Slightly tedious, but elementary manipulations, yield the result, 
$(\Delta b)^2=A^{-2}(n+1)(n+2)^2(n+3)/8$.

This gives $\Delta b/b=\sqrt{8}/[(n+2)\sqrt{(n+1)(n+3)}] \propto A^{-2/3}$.
In systems with $\simeq 10^3$ particles a ``horizontal'' bar of order $1\%$
might likely be neglected, but many systems studied in atomic, molecular or
nuclear physics rather deal with a few scores of particles or hardly two or
three hundred of them, and a horizontal uncertainty ranging between
$\simeq 8\%$ and $\simeq 2\%$ might trigger some attention.

\subsection{Fluctuation of the quadrupole moment in a deformed harmonic
oscillator shell model}

The Hamiltonian now reads, $H=\sum_i [p_i^2+x_i^2+y_i^2+z_i^2/\gamma^4]/2$,
hence prolate shapes if $\gamma>1$ and oblate ones if $\gamma<1$. The 
collective operator is now chosen as, 
$B=\sum_i \beta_i$ with $\beta_i=2z_i^2-x_i^2-y_i^2$, without a precaution
cut-off. For values of $\gamma$ close enough to $1$ to induce a weak
splitting of levels we can study the same scheme of fully filled shells as
for the model just above. Orbitals $\varphi_i$ in a ``shell $m$'' have wave
functions $\chi_{m^{\prime}}(x)\, \chi_{m^{\prime \prime}}(y)\,
\chi_{m^{\prime \prime \prime}}(z/\gamma)/ \sqrt{\gamma}$, where the 
$\chi$'s are the standard 1-D harmonic oscillator states and 
$m^{\prime}+m^{\prime \prime}+m^{\prime \prime \prime}=m$. The orbital
energies are, 
$m^{\prime}+m^{\prime \prime}+1+(m^{\prime \prime \prime}+1/2)/\gamma^2$,
obviously.

When all the orbitals $\varphi_i$ of the shells with labels $0,1,\dots,n$
are filled, the particle number for spinless and isopinless fermions is again,
$A=(n+1)(n+2)(n+3)/6$ and the value of the quadrupole is, obviously, 
$b \equiv \langle B \rangle = \sum_{i=1}^A \langle \varphi_i | \beta |
\varphi_i \rangle$. Its quantum fluctuation is given by the same 
particle-hole summation, 
$(\Delta b)^2=\sum_{Ph} (\langle P | \beta | h \rangle)^2$. This also reads,
$(\Delta b)^2={\rm Tr}\, \beta\, (1-\sigma)\, \beta\, \sigma$, where we use
the one-body density operator,
$\sigma=\sum_i^A |\varphi_i \rangle \langle \varphi_i |$, namely the projector
upon holes. Its complement, $(1-\sigma)$, is the projector upon the particle
subspace. The trace, ${\rm Tr}$, runs in one-body space. This yields,
$(\Delta b)^2={\rm Tr}\, \beta^2\, \sigma-{\rm Tr}\, \beta\, \sigma\, \beta\,
\sigma$. Since $\sigma$ is here real symmetric and $\beta$ is a local
operator, the first and second terms read, in coordinate representation, 
$\int d\vec r\, \sigma(\vec r,\vec r)\, [\beta(\vec r)]^2$, and,
$-\int d\vec r\, d\vec r^{\, \prime}\, \beta(\vec r)\, 
[\sigma(\vec r,\vec r^{\, \prime})]^2\, \beta(\vec r^{\, \prime})$,
respectively. The symbol $\vec r$ is here a short notation for the three
coordinates $x,y,z$.

Quadrupole values,
$b=(\gamma^2-1) (n+1)(n+2)^2(n+3)/12$, are easily found,
with $n$ the label of the highest filled ``shell'',  and, accordingly,
the particle number, $A=(n+1)(n+2)(n+3)/6$.
Because of the factor, $(\gamma^2-1)$, such values of the
quadrupole do vanish when $\gamma=1$, as expected for spherical shells, and
also show the correct signs for prolateness and oblateness.

Similar brute force considerations provide $(\Delta b)^2= 
(1+2 \gamma^4) (n+1)(n+2)^2(n+3)/12$.
It would not  be significant to discuss the ratio, $\Delta b/|b|$, 
when $\gamma$ is close to $1$, obviously, but for a value of $\gamma$ such
as $3/2$, for instance, the ``full shell'' particle numbers considered above
return the following sequence of values for this relative uncertainty, 
2.7,\, 1.1,\, 0.60,\, 0.38,\, 0.26,\, 0.19,\, 0.15,\, 0.11,\, 0.09,\, 0.08,
\, 0.06. If $\gamma=1/2,$ the sequence becomes, 
1.4,\, 0.58,\, 0.32,\, 0.20,\, 0.14,\, 0.10,\, 0.08,\, 0.06,\, 0.05,
\, 0.04,\, 0.03. Both sequences show that the relative uncertainty shrinks
when the particle number increases, as should be expected, but, as long as
the particle number does not exceed a few hundred, fluctuations of 
$\simeq 5 \% $ cannot be neglected.

\begin{figure}[h]
\centering\scalebox{0.68}{\includegraphics*{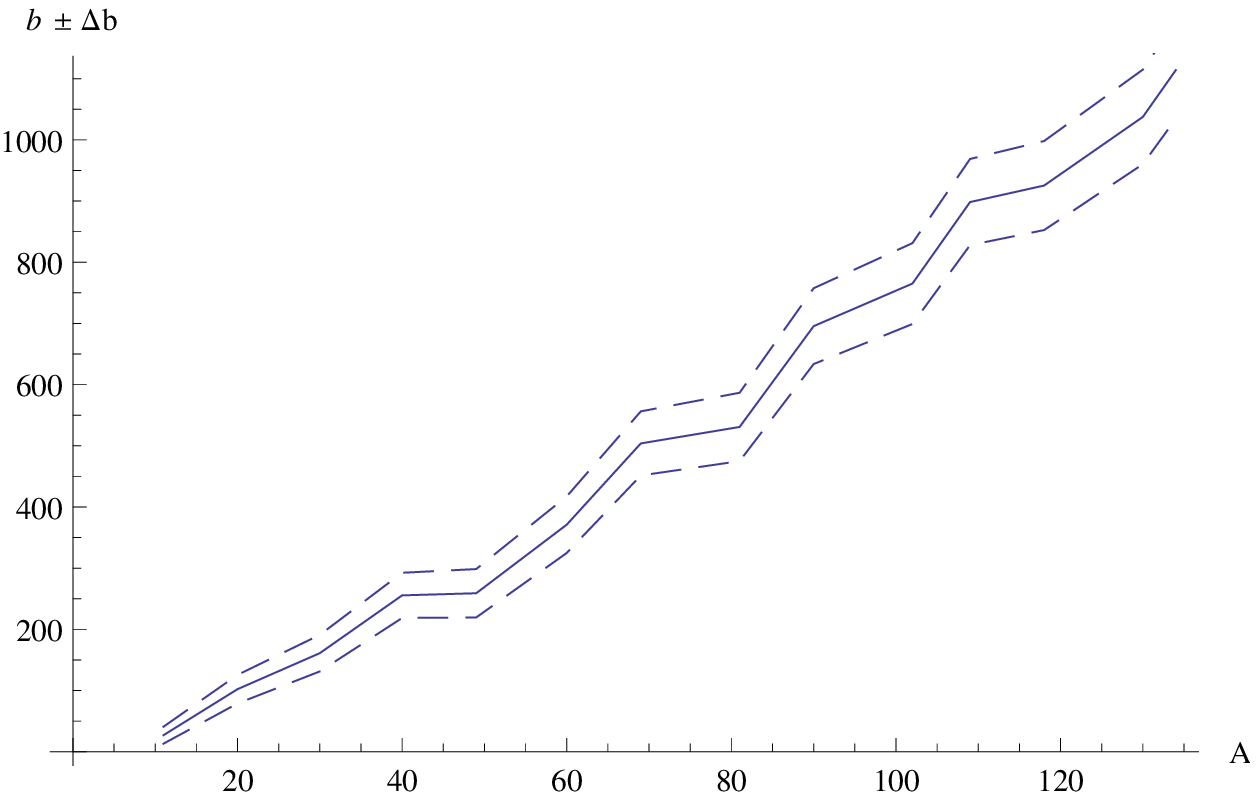}}
\caption{\label{figure1} Quadrupole values $b$ (full line) and 
$b \pm \Delta b$ (dashed lines) for $\gamma=4/3$.}
\end{figure}

\begin{table}[h]
\begin{center}
\begin{tabular}{|c|c|c|c|}
\hline
A& Quadrupole& Square deviation& Relative uncertainty \\
\hline
11& 27& 575/3& 0.51 \\
\hline
20&  920/9& 44200/81& 0.23\\
\hline
30& 1450/9&  72230/81& 0.19\\
\hline
40& 2300/9&  110500/81& 0.14\\
\hline
49& 259& 14069/9& 0.15\\
\hline
60& 3340/9&  173300/81& 0.13\\
\hline
69& 4535/9& 222241/81& 0.10\\
\hline
81&  4777/9&  252863/81& 0.11\\
\hline
90& 6260/9& 311020/81& 0.089\\
\hline
102& 6886/9& 353930/81& 0.086\\
\hline
109& 2695/3& 133841/27& 0.078\\
\hline
118&  2776/3& 142952/27& 0.079\\
\hline
 130& 9338/9&  484054/81& 0.075\\
\hline
\end{tabular}
\caption{Typical results for $\gamma=4/3$.}
\end{center}
\end{table}
Actually, level crossing occurs when deformation sets in. Full 
fillings of previously spherical shells do not represent ground states of
deformed Hamiltonians. 
Set $\gamma=4/3$ for instance, and, given a particle number $A$, fill the 
lowest orbitals of the corresponding deformation scheme. This gives quadrupole,
mean square deviations and relative uncertainty values listed in Table I. In Figure~\ref{figure1}
we show, in terms of $A$, quadrupole values $b$ along a full line and values 
$b \pm \Delta b$ along dashed lines. 

\begin{figure}[b]
\centering\scalebox{0.68}{\includegraphics*{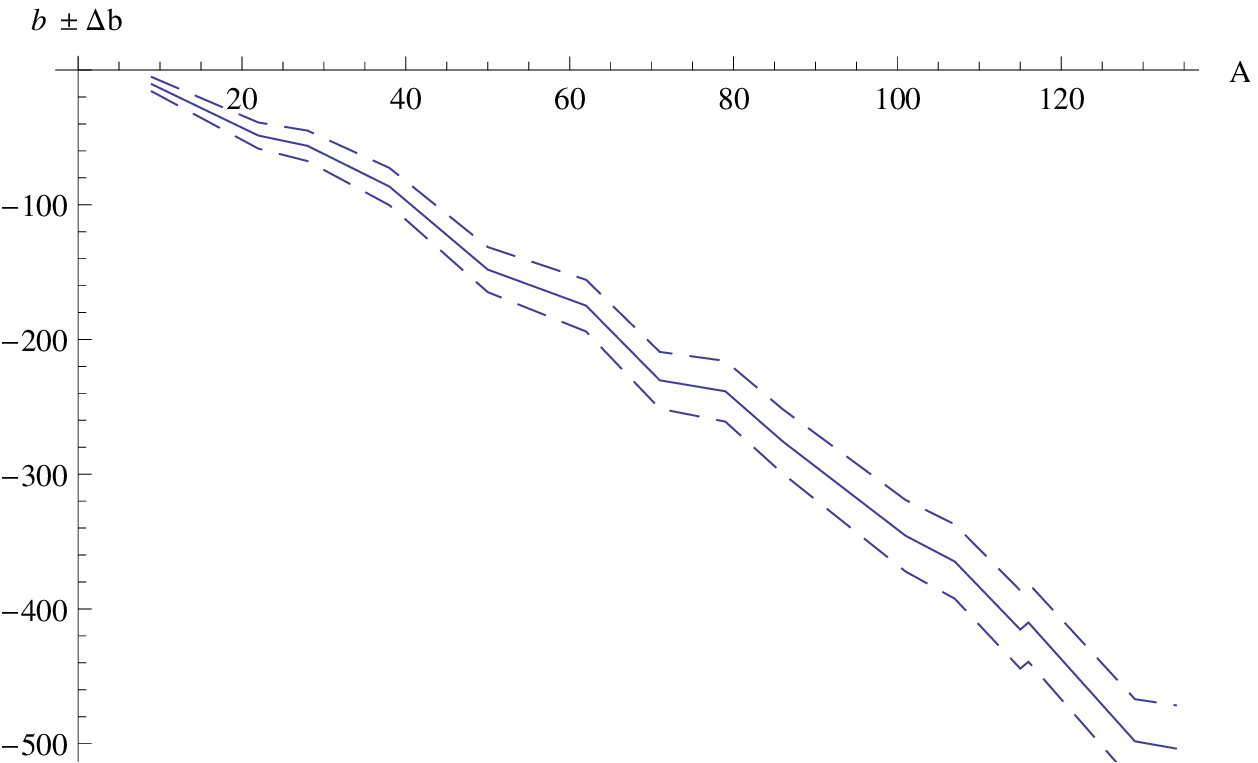}}
\caption{\label{figure2} Quadrupole values $b$ (full line) and 
$b \pm \Delta b$ (dashed lines) for $\gamma=3/4$.}
\end{figure}

With $\gamma$=3/4, typical results are listed in Table II and shown in Fig.~\ref{figure2}.
Relative fluctuations $\Delta b/|b|$ decrease when $A$ increases, as should
be expected, but an order of magnitude of several percent at least cannot be
avoided.
 
\begin{table}[h]
\begin{center}
\begin{tabular}{|c|c|c|c|}
\hline
A& Quadrupole& Square deviation& Relative uncertainty \\
\hline
 9&  -169/16&    3647/128& 0.51\\
\hline
22&  -389/8&      6019/64& 0.20\\
\hline
28&  -225/4&      4095/32& 0.20\\
\hline
38&  -173/2&      3067/16& 0.16\\
\hline
50&  -1185/8&    17895/64& 0.11\\
\hline
62&  -1399/8&    23585/64& 0.11\\
\hline
71&  -3685/16&  56867/128& 0.092\\
\hline
79&  -3815/16&  64945/128& 0.094\\
\hline
86&  -2205/8&    36459/64& 0.087\\
\hline
101&  -5531/16&  90301/128& 0.077\\
\hline
107&  -5837/16&  97339/128& 0.076\\
\hline
116&  -3281/8&    54247/64& 0.071\\
\hline
129&  -7973/16& 125443/128& 0.063\\
\hline
\end{tabular}
\caption{Typical results for $\gamma=3/4$.}
\end{center}
\end{table}

If spin, for electrons, and both spin and isospin are reinstated, for nucleons,
keeping the same occupied levels while particle number is multiplied by $2$
and $4$, respectively, it is obvious that $\Delta b/|b|$ is divided by 
$\sqrt{2}$ and by $2$, respectively. Actually, for nucleons, the proton number
is usually smaller than that of neutrons, and orbitals may somewhat differ,
hence the exact reduction factor of $\Delta b/|b|$ will slightly differ from
$2$, but this changes nothing to the fact that quantum fluctuations exist and
a relative uncertainty of at least a few percent cannot be avoided.

\subsection{Fluctuation in a constrained diagonalization}

\begin{figure}
\centering\scalebox{0.68}{\includegraphics*{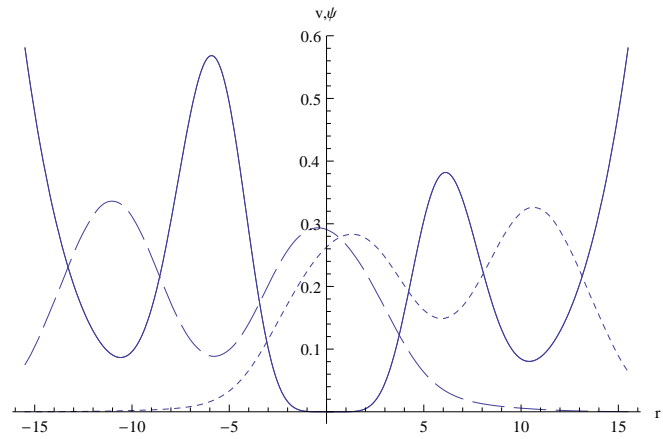}}
\caption{\label{figure3} Toy model, 1-D potential, full curve. 
Tunnel effect of constrained eigenstate under left barrier, dashed curve. 
Same effect under right barrier, dotted curve.}
\end{figure}

We consider in this section a one-dimentsional Hamiltonian, $H = -d^2/dr^2+v(r)$, with a double hump
potential, 
$v=(r-1/5)^6/5000\, e^{-r^2/12} + (r/12)^6/8$, shown as a full line in 
Fig. \ref{figure3}.

\begin{figure}
\centering\scalebox{0.68}{\includegraphics*{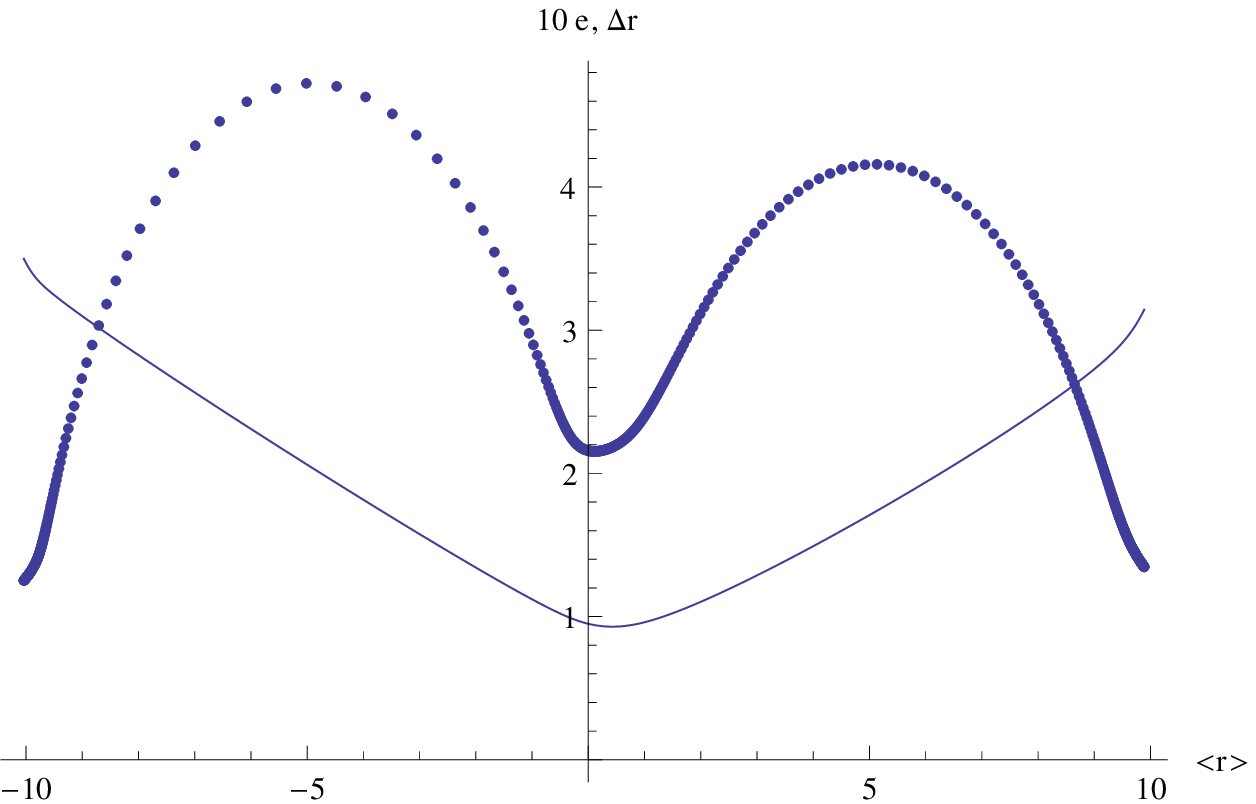}}
\caption{\label{figure4} Same toy model. Energy curve, full line. Fluctuation
curve, dots. Strong increase of $\Delta r$ when the constrained eigenstate
delocalizes into two wave packets.}
\end{figure}

We take the constrained
Hamiltonian as ${\cal H}=H-\lambda\, r$ and diagonalize it
in a subspace spanned by shifted Gaussians. While the eigenstate,
$\psi_{\lambda}(r),$ shows a single wave-packet when the average value, 
$\left\langle r \right\rangle \equiv \left\langle \psi_{\lambda} \left| r 
\right| \psi_{\lambda} \right\rangle,$ sits near a minimum of $v$, an expected
tunnel effect occurs when $\left\langle r \right\rangle$ sits near a maximum
of $v$. There $\psi$ shows two connected packets, one at each side of the
barrier, inducing a lowering of the energy. Such a bimodal (even multimodal
in several extreme cases we tested) situation induces a very bad probing of the
barrier. The plot of the energy, $e=\langle H \rangle$, 
(multiplied ten times for graphical reasons) in terms of
$\langle r \rangle$ does not reflect $v(r$) in any way, see the thin
full line in Fig. \ref{figure4}. When tunnel effects occur, fluctuations, 
$\Delta r = \sqrt{\left\langle r^2 \right\rangle - \left\langle r 
\right\rangle^2},$
are dramatically larger than when a unique packet sits in a valley. The
label, $\langle r \rangle$ in our case, is thus misleading. Although
$\mathcal{H}$ was exactly diagonalized, constrained variation generated a bad
quality representation of $v(r)$. Fig. \ref{figure4} illustrates how big the
uncertainty on $\langle r \rangle$ can become. \textit{Moreover, 
this ``energy surface'' turns out to be convex.}

The previous models in this Section show that somewhat large quantum
fluctuations of collective coordinates do exist, but the present model, in
this Subsection, illustrates a new fact, namely that such fluctuations may
vary along an energy surface and may influence the surface itself. 
A trivial way to prevent
fluctuations from arbitrarily varying is to introduce a double constraint
via the square, $B^2$, of the initial constraint 
operator \cite{PhysScrip}. One adjusts the second Lagrange multiplier so that the fluctuation, 
$\Delta b$, remains small, and, for a stable quality
of the representation, reasonably constant. (Alternately one can tune the
second Lagrange multiplier to ensure more or less constant and/or small
enough ratios $\Delta b/|b|$.)
Again with 1-D toy Hamiltonians of the form, 
${\cal H}'=-d^2/dr^2+v(r)-\lambda_1 r+\lambda_2 r^2,$
or, equivalently, ${\cal H}'=-d^2/dr^2+v(r)+\lambda_2(r-\lambda_1)^2,$
we tuned $\lambda_2$ into a function $\lambda_2(\lambda_1)$ to enforce
a unimodal situation  with $\Delta r$ kept constant when $\langle r \rangle$
 evolves. A sharp, stable probe of the barrier results. Convexity is then 
``defeated''. The true shape of $v$ is recovered \cite{PhysScrip}.

\section{Discussion and conclusion}

The present work shows that expectation values of collective
coordinate operators may lead to misleading coordinates for an energy
surface. Convexity is a major property of any energy surface obtained by 
an exact minimization of the energy under constraint(s), and may occur
through tunnel effects and collective coordinate fluctuations. If an energy
landscape with ``saddles'' is needed, such deviations from convexity
contradict the requirement that the energy must transit through exact minima. The
success of collective
models that use a non-trivial landscape via mean field approximations 
is too strong to be rejected as 
physically and/or mathematically unsound, but its validation likely relates
to further methods such as, resonating group methods \cite{WilTan}, generator 
coordinate (GC) ones \cite{HilGriWhe,HilGriWhea}, Born-Oppenheimer approximations,
influence functionals \cite{FeyVer}, deconvolutions of wave packets in
collective coordinate spaces \cite{GirGra}, etc. In particular, one can
argue that, while individual exact states $\psi_{\lambda}$ or
mean-field ones $\phi_{\lambda}$ may carry large uncertainties for the label
$b,$ such states may still provide a good global set for a GC calculation.
But then the physics lies as much in non-diagonal elements $H(b,b')$ and 
$N(b,b')$ of the GC energy and overlap kernels, respectively, than in the
diagonal, $e(b) \equiv H(b,b).$

Even so, except for anharmonic vibrations, where one valley is sufficient,
one has to justify why concave branches can be as significant as branches
obtained from exact minimizations of the energy.

Because of kinetic terms, which enforce delocalizations, a full Hamiltonian is
often not well suited to ensure a good localization of the operators, $B$, a necessary 
condition for the exploration of an energy surface parametrized by their expectation values, $b$.
Recall that, in the Born-Oppenheimer treatment of the hydrogen molecule,
the proton kinetic energy operator is initially removed from the Hamiltonian,
allowing the collective coordinate, namely the interproton distance, to be 
frozen as a zero-width parameter. Most often in nuclear, atomic, and molecular
physics, such a removal is not available. Hence, such an approach may lead to ``dangerous consequences'',
as illustrated by Fig. \ref{figure4}, from the toy Hamiltonian
introduced at the stage of Fig. \ref{figure3}. The growth of fluctuations due
to tunnel effects is spectacular. Fortunately, if one forces the constrained
eigenstate to retain a narrow width while the collective label runs, convexity 
is defeated at the profit of a reconstruction of the potential shape.

The technical devices used with success in
Refs. \cite{GLW,FQVVK,FQVVKa,SSBN} require a further investigation of the solutions they 
generated. Uncertainties in the collective coordinates must be acknowledged. ``Mountains'' can be 
underestimated, as is the case when tunnel effects deplete the energy probing wave 
function. (A hunt for multimodal solutions would therefore be useful.) 
Since fluctuations are important at ``phase transitions'', collective
operators must be completed by their own squares, in combinations of the form, 
$\mathcal{K}=H-\lambda B + \mu(\lambda)  B^2$,
with $\mu(\lambda)$ adjusted to avoid wild increases of $\Delta b$ . Such 
operators $\mathcal{K}$ govern both a constraint and its fluctuation, but 
obviously differ from a double constraint form,
$\mathcal{H}=H-\lambda_1 B+\lambda_2 B^2$, with two independent parameters,
$\lambda_1,\lambda_2$. The second derivative, $d^2 \varepsilon/d\lambda^2$
contains additional terms due to $d\mu/d\lambda$ and $d^2\mu/d\lambda^2$,
hence a one-dimensional path with non convex structures can be induced by
$\mathcal{K}$ inside that convex two-dimensional landscape due to $\mathcal{H}$.
We verified this ``adjustable $B^2$'' method with many numerical tests.

In theories using, partly at least, liquid drop models, see Ref. \cite{Sierk} for
instance, labels $b$ are purely classical. Such theories are thus safe from
the present concerns. But many other energy surfaces used in realistic
situations come from mean-field constrained calculations. It remains to be
tested whether their solutions, stable or metastable, carry a mechanism that
diminishes the fluctuation of collective degrees of freedom. This mechanism,
if it exists, deserves investigation. We conclude that a review of landscapes
obtained by constrained HF or HB is in order, to analyze the role of
collective coordinate fluctuations. It is clear that such surfaces deserve
corrections because of likely variable widths $\Delta b$ of their collective
observables, and also, obviously, because convolution effects and zero-point
energies must be subtracted.

To summarize, we first found that collective operators carry significant
fluctuations. We also found that fluctuations, and convexity situations,
can enforce poor energy landscapes if constrained energy minimizations are
used with fixed operators $B$ and fixed trial spaces. Unacceptable uncertainties, $\Delta b$,
can vitiate the meaning of collective labels. We did even 
discover bi- or multimodality in mean field approximations. Fortunately,
given the same fixed operators and trial spaces, a modest deviation from
fixed constraints, namely adjustable combinations of $B$ and $B^2,$ commuting 
operators indeed, allows an analysis ``at fluctuations under control'', 
with  unimodal probes of landscapes and a controlled quality of the 
collective representation. A puzzling question remains: is the good quality
of constrained mean field solutions in the literature \cite{FQVVK,FQVVKa} the 
result of a ``self damping'' of collective coordinate fluctuations? Are 
multimodal situations blocked when nuclear mass increases?

SK acknowledges support from the National Research Foundation of South Africa.
BG thanks N. Auerbach for stimulating discussions and the University of
Johannesburg  and the Sackler Visiting Chair at the University of Tel Aviv 
for their hospitality during part of this work.

\section*{References}
\bibliography{flu_sk}

\end{document}